\documentclass[12pt,preprint]{aastex}
\usepackage{color}
\usepackage{amsmath}
\shorttitle{A BOUND-VIOLATING GROUP}
\shortauthors{Lee et al.}
\begin{document}
\title{Linear Tidal Vestige found in the WM Sheet}
\author{Jounghun Lee\altaffilmark{1}, 
Suk Kim\altaffilmark{2}, and Soo-Chang Rey\altaffilmark{3}}
\altaffiltext{1}{Astronomy Program, Department of Physics and Astronomy, Seoul National University, 
Seoul 151-742, Korea
\email{ jounghun@astro.snu.ac.kr}}
\altaffiltext{2}{Korea Astronomy \& Space Science institute, 776 Daedeokdae- ro, Daejeon 305-348, Korea}
\altaffiltext{3}{Department of Astronomy and Space Science, Chungnam National University,Daejeon 305-764, Korea}
\begin{abstract} 
We present a vestige of the linear tidal influence on the spin orientations of the constituent galaxies of the 
WM sheet discovered in the vicinity of the Virgo cluster and the Local Void. The WM sheet is chosen as an optimal target 
since it has a rectangular parallelepiped-like shape whose three sides are in parallel with the supergalactic Cartesian axes.  
Determining three probability density functions of the absolute values of the supergalactic Cartesian components of the spin 
vectors of the WM sheet galaxies, we investigate their alignments with the principal directions of the surrounding large-scale 
tidal field. When the WM sheet galaxies located in the central region within the distance of $2\,h^{-1}$Mpc are excluded,   
the spin vectors of the remaining WM sheet galaxies are found to be weakly aligned, strongly aligned, and strongly anti-aligned 
with the minor, intermediate and major principal directions of the surrounding large-scale tidal field, respectively. 
To examine whether or not the origin of the observed alignment tendency from the WM sheet is the linear tidal effect, 
we infer the eigenvalues of the linear tidal tensor from the axial ratios of the WM sheet with the help of the 
Zel'dovich approximation and conduct a full analytic evaluation of the prediction of the linear tidal torque model for 
the three probability density functions. A detailed comparison between the analytical and the observational results 
reveals a good quantitative agreement not only in the behaviors but also in the amplitudes of the three probability 
density functions. 
\end{abstract}
\section{Introduction}\label{sec:intro}

The anisotropic large-scale tidal field has an effect of producing a web-like pattern in the spatial distribution of the 
galaxies, which is often collectively referred to as the cosmic web \citep{web96}. The voids, filaments, sheets and knots 
observed on the largest scales are the four constituent elements of the cosmic web and called the large-scale 
structures, whose formations are closely linked to the coherence of the principal directions of the 
large-scale tidal field \citep[e.g.,][]{hah-etal07,for-etal09,cau-etal13}. 
The three dimensional stretch (compression) of matter along three principal directions of the coherent large-scale tidal 
fields lead to the formation of voids (knots).  
The sheets (filaments) form at the boundaries between the voids through one (two) dimensional compression 
of matter only along the major (along the major and intermediate) principal directions \citep{zel70}. 
The largest knots corresponding to the rich galaxy clusters in the universe often form at the intersections of the 
filaments along which the galaxies anisotropically merge with one another \citep{wes-etal95}, 
while the smaller knots are mostly distributed along the filaments or embedded in the sheets. 

Other than weaving the cosmic web, the linear tidal field has an effect of originating the spin
angular momenta of the proto-galaxies. The linear tidal torque (LTT) model developed by \citet{dor70} 
explains in the frame of the first-order Lagrangian perturbation theory that the exertion of the torque force of the 
linear tidal field causes an asymmetric proto-galactic region to possess the spin angular momentum 
\citep[see also][]{whi84}.  
The signature prediction of this LTT model is that the spin axes of the proto-galaxies tend to be strongly aligned, weakly  
aligned and strong antialigned with the intermediate, minor and major principal axes of the linear tidal tensors, respectively  
\citep{LP00,LP01,por-etal02}. In this model, the coherence in the principal directions of the 
linear tidal field is translated into that in the spin directions of the proto-galaxies. 

Provided the linear tidal field can be well approximated by the large scale tidal field whose coherence is responsible for the 
formation of the cosmic web and that the spin axes of the present galaxies preserve well their initially induced tidal alignments, 
the effect of the linear tidal field on the spin directions of the proto-galaxies would be reflected at least to some degree 
by the alignments of the spin axes of the present galaxies with the shapes of the large-scale structures, since the eigenvectors 
of the inertia momentum tensors of the large-scale structures coincide well with the principal axes of the large-scale tidal field  
\citep{LP00,por-etal02}. 

However, the spin directions of the galaxies after the turn-around moments are prone to severe nonlinear modifications 
especially during the compression and/or stretch of the surrounding matter distribution to form the large-scale structures. 
The N-body experiment performed by \citet{che-etal16} demonstrated that only for the case of those large-scale 
structures that are still in the quasi-linear regime,  it is possible to find a vestige of the linear tidal effect from 
the spin directions of the embedded galaxies.  
Another recent numerical study done by \citet{bri-etal16} showed that the galaxies embedded in the sheets have 
relatively low velocity dispersions compared with those in the filaments, which implied that the sheets are still in the 
quasi-linear regime, which is consistent with the numerical finding that the first objects ever condensed out are likely 
to have sheet-like shapes \citep{sha-etal95}. 
Given these numerical results, we suggest that the galaxies embedded in the sheets are the 
best targets for the detection of the linear tidal effect on their spin axes.

Another merit of using the sheets is that the direction of at least the major principal axis of the surrounding large-scale 
tidal field can be relatively accurately determined \citep{nav-etal04}. For the case of a flat sheet, the major principal axis of the 
tidal field should be in the direction normal to the face of the sheet (called the sheet normal vector). For the case of a sheet 
wrapping a void, the radial vector from the center of a given void to a galaxy position in the void surface (void radial vector) 
must be in the direction of the major principal axis of the surrounding large-scale field. If a sheet truly preserved the initially 
imprinted memory of the linear tidal alignment at a detectable level, then the spin directions of the galaxies embedded in the 
sheet would show a tendency of being anti-aligned with the sheet normal vector (or equivalently, a tendency to lie in the 
plane of the sheet). 

Several observational efforts were already made to dig up the linear tidal vestige from the sheet-like structures. 
For instance, \citet{nav-etal04} claimed that the anti-alignments of the spin axes of the local spiral galaxies 
belonging to the Local Supercluster with the direction of the supergalactic north pole should be taken as an evidence 
for the presence of the linear tidal vestige. 
\citet{tru-etal06} investigated whether or not the spin axes of the spiral galaxies on the surfaces of the voids 
were anti-aligned with the void radial directions and claimed a detection of statistically significant signal. 
\citet{hir-etal17} noted that the origin and strength of the intrinsic spin alignments of the galaxies depend on 
the Hubble types of the galaxies, and showed that for the case of the Sab spiral galaxies located 
in the sheets, their spin directions show a tendency of anti-alignments with the sheet normal vector. 
These observational findings seemed quite consistent qualitatively with the trend predicted by the LTT model.

Yet, the presence of the linear tidal vestige in the sheets are still inconclusive, having been challenged by the 
recent reports of the counter evidences. \citet{SW09} reinvestigated the spin directions of the galaxies at the void 
surfaces and found no signal of their anti-alignments with the  void radial directions for the case of 
the spherical voids. \citet{var-etal12} confirmed the finding of \citet{SW09} with a larger sample and claimed 
that the alignment tendencies show variation with the sizes of the voids.  
A similar counter evidence was also provided by \citet{TL13} who found no preferred directions in the orientations 
of the spin axes of the sheet galaxies, identifying the sheets using the "Biscous model". 
Given many numerical results that the spin axes of the galactic halos embedded in the sheets tend to be 
anti-aligned with the sheet normal vectors  \citep[e.g.,][]{ara-etal07,cue-etal08,lib-etal13,for-etal14}, 
those observational counter evidences may be interpreted not as real challenges but as an indication of 
the presence of systematics such as inaccurate measurements of the spin directions of the void galaxies, 
ambiguity and uncertainty involved with the identification of the voids and sheets, and so forth. 

However, the previous works were inconclusive not only about the detection of the signal itself but also about 
the origin of the signal. The subsequent nonlinear evolution of the galaxies after the turn-around moments 
would not necessarily weaken but rather could strengthen the inclination of their spin axes to be aligned or anti-aligned 
with the shapes of the surrounding large-scale structures \citep{cod-etal12,dub-etal14,wel-etal14,lai-etal15}. 
To distinguish between the spin alignments originated by the linear tidal influences and those 
generated/strengthened by the nonlinear processes, what is required is an exact evaluation of the prediction 
of the LTT model on the spin alignments of the sheet galaxies and a detailed comparison of an observationally detected 
signal of the spin alignments with the prediction of the LTT model as well.  

The linear tidal effect cannot produce an arbitrary high signal of the galaxy spin alignments but rather limits the strength 
of the signal. Only when the strength of an alignment signal from the galaxies embedded in a sheet is found to lie in the 
range constrained by the LTT model and the alignment tendencies agree quantitatively with the signature prediction 
of the LTT model as mentioned in the above,  it can be declared as a detection of the linear tidal vestige from the sheet. 
In this paper, we aim at detecting the linear tidal vestige from a newly discovered sheet with a conspicuously rectangular 
parallelepiped shape.

We organize this paper to present the followings: A review of the LTT model and a full analytic derivation of its prediction 
on the spin directions of the sheet galaxies in Section \ref{sec:review}.  The results of testing the predictions of the LTT 
model against new observational data in Section \ref{sec:test}.  The summary of the final result and discussion of its 
implication as well as two caveats in Section \ref{sec:sum}. 

\section{A Review of the Linear Tidal Torque Model}\label{sec:review}

\subsection{Probability distributions of the unit galaxy spins}

Suppose that a proto-galaxy located at a position ${\bf r}$ has an angular momentum, ${\bf J}$, and that at the 
same position the linear tidal field ${\bf T}=(T_{ij}) $ smoothed on the galactic mass scale has three distinct eigenvalues,  
$\{\lambda_{1}, \lambda_{2}, \lambda_{3}\vert \lambda_{1}\ge \lambda_{2}\ge\lambda_{3}\}$. 
The linear tidal torque (LTT) theory relates ${\bf J}$ to ${\bf T}$ by modeling the conditional probability density 
function of ${\bf J}$ given ${\bf T}$ as the following multivariate Gaussian distribution \citep{CT96,LP01,por-etal02}
\begin{equation}
\label{eqn:pro_j}
p({\bf j}\vert {\bf T}) = p(J,{\bf j}\vert{\bf T})= \frac{1}{[(2\pi)^3 {\rm det}({\bf U})]^{1/2}}
\exp\left[-\frac{J^{2}}{2}\left({\bf j}\cdot{\bf U}^{-1}\cdot{\bf j}\right)\right]\, .
\end{equation}
Here ${\bf j}$ denotes the unit spin vector of a proto-galaxy defined as 
${\bf j}\equiv {\bf J}/\vert{\bf J}\vert$, and ${\rm det}\left({\bf U}\right)$ is the determinant of the covariance matrix 
${\bf U}=(U_{ij})\equiv \langle j_{i}j_{j}\vert{\bf T}\rangle$ whose components are given as  \citep{LP00,LP01}
\begin{equation}
\label{eqn:cov}
U_{ij} = \frac{1+c_{t}}{3}\delta_{ij} - c_{t}\sum_{k=1}^{3}{\hat{T}_{ik}\hat{T}_{kj}}\, ,
\end{equation}
where $\hat{\bf T}$ is the unit traceless tidal shear tensor defined as 
$\hat{\bf T} \equiv \tilde{\bf T}/\vert\tilde{\bf T}\vert$ with $\tilde{\bf T}\equiv {\bf T}-(1/3)\sum_{i=1}^{3}{T_{ii}}$, 
and $c_{t}$ is a stochastic tidal link parameter that measures the strength of the correlation between ${\bf j}$ and ${\bf T}$. 
The value of $c_{t}$ depends on the degree of the misalignments between the principal axes of ${\bf T}$ and those of the 
inertia momentum tensor, ${\bf I}$, of a proto-galaxy.  The more misaligned they are with one another, the closer to unity 
the value of $c_{t}$ can have \citep{LP00,por-etal02}. 

Let $\hat{\bf p}_{1},\ \hat{\bf p}_{2}$ and $\hat{\bf p}_{3}$ denote the major, intermediate and minor principal axes of 
${\bf T}$, respectively (i.e., the orthonormal eigenvectors corresponding to $\lambda_{1},\ \lambda_{2}$ and 
$\lambda_{3}$, respectively). In the principal axis frame of ${\bf T}$, the conditional probability density function of ${\bf j}$ 
can be obtained by marginalizing Equation (\ref{eqn:pro_j}) over $J\equiv \vert{\bf J}\vert$ \citep{LP01}:  
\begin{eqnarray}
\label{eqn:pro_hj}
p({\bf j}\vert{\bf T})&=&\int\,p({\bf j}\vert{\bf T}) J^{2}dJ \nonumber \\
&=& \frac{1}{2\pi}\left[\left(1+c_{t}-3c_{t}\hat{\lambda}^{2}_{1}\right)
\left(1+c_{t}-3c_{t}\hat{\lambda}^{2}_{2}\right)
\left(1+c_{t}-3c_{t}\hat{\lambda}^{2}_{3}\right)\right]^{-\frac{1}{2}}\times \nonumber \\
&&\left(\frac{\vert\hat{\bf p}_{1}\cdot
{\bf j}\vert}{1+c_{t}-3c_{t}\hat{\lambda}^{2}_{1}} + 
\frac{\vert\hat{\bf p}_{2}\cdot{\bf j}\vert}{1+c_{t}-3c_{t}\hat{\lambda}^{2}_{2}} + 
\frac{\vert\hat{\bf p}_{3}\cdot{\bf j}\vert}{1+c_{t}-3c_{t}\hat{\lambda}^{2}_{3}}\right)^{-\frac{3}{2}}\, , 
\end{eqnarray}
where $\hat{\lambda}_{1},\ \lambda_{2},\ \lambda_{3}$ are the eigenvalues of $\hat{\bf T}$ distributed over a 
sphere with a radius of unity, satisfying the constraints of  
$\sum_{i=1}^{3}{\hat{\lambda}^{2}_{i}}=1$ and $\sum_{i=1}^{3}{\hat{\lambda}_{i}}=0$ \citep{LP01}. 

The conditional probability density of the absolute value of the ${\bf p}_{i}$-axis component of ${\bf j}$ 
(i.e, $\vert\hat{\bf p}_{i}\cdot {\bf j}\vert$) can be obtained as
\begin{equation}
\label{eqn:pro_hjp}
p(\vert\hat{\bf p}_{1}\cdot{\bf j}\vert)=\int_{0}^{2\pi}p({\bf j})\,d\phi_{23}, \quad
p(\vert\hat{\bf p}_{2}\cdot{\bf j}\vert)=\int_{0}^{2\pi}p({\bf j})\,d\phi_{31}, \quad
p(\vert\hat{\bf p}_{3}\cdot{\bf j}\vert)=\int_{0}^{2\pi}p({\bf j})\,d\phi_{12}, 
\end{equation}
where 
\begin{equation}
\label{eqn:phi}
\phi_{23}\equiv \tan^{-1}
\left(\frac{\hat{\bf p}_{2}\cdot {\bf j}}{\hat{\bf p}_{3}\cdot {\bf j}}\right)\, ,\quad 
\phi_{31}\equiv \tan^{-1}
\left(\frac{\hat{\bf p}_{3}\cdot {\bf j}}{\hat{\bf p}_{1}\cdot {\bf j}}\right)\, ,\quad 
\phi_{12}\equiv \tan^{-1}
\left(\frac{\hat{\bf p}_{1}\cdot {\bf j}}{\hat{\bf p}_{2}\cdot {\bf j}}\right)\, ,
\end{equation}

Equation (\ref{eqn:pro_hj}) is not uniform provided that 
$c_{t}\ne 0$ and $\lambda_{i}\ne \lambda_{j}$ for $i, j\in \{1,2, 3\}$.
Depending on the sign of $c_{t}$, $p({\bf j})$ behaves differently: for the case of $c_{t}>0$, it yields 
\begin{equation}
\label{eqn:ctg0}
\langle\vert\hat{\bf p}_{2}\cdot{\bf j}\vert\rangle > 0.5\sim 
\langle\vert\hat{\bf p}_{3}\cdot{\bf j}\vert\rangle> \langle\vert\hat{\bf p}_{1}\cdot{\bf j}\vert\rangle\, ,
\end{equation}
which implies that the spin vectors of the proto-galaxies are strongly and weakly aligned with $\hat{\bf p}_{2}$ and 
$\hat{\bf p}_{3}$, respectively, but strongly anti-aligned with $\hat{\bf p}_{1}$. Whereas, for the case of $c_{t}<0$, 
the inequality sign is reversed as
\begin{equation}
\label{eqn:ctl0}
\langle\vert\hat{\bf p}_{2}\cdot{\bf j}\vert\rangle < 
\langle\vert\hat{\bf p}_{3}\cdot{\bf j}\vert\rangle\sim 0.5 < \langle\vert\hat{\bf p}_{1}\cdot{\bf j}\vert\rangle.
\end{equation} 

The above unique prediction of the LTT model on the alignments between ${\bf j}$ and ${\bf T}$ is valid in the linear regime 
where the proto-galactic regions remain coupled to the surrounding linear tidal field before their turn-around moments.  
In the non-linear regime, the tidal field would develop non-Gaussianity on the galactic mass scale and the spin directions 
of the present galaxies would inevitably deviate from those of their progenitors.  However, if some galaxies did not undergo 
severe nonlinear modifications in their spin orientations, the linear tidal influence on their spin directions would be manifested 
by their tendency of being aligned with the principal axes of the present large-scale tidal fields that can approximate the 
linear tidal field in the proto-galactic stage.  In other words, Equations (\ref{eqn:pro_hjp}) and Equation (\ref{eqn:ctl0}) could be 
used to describe the alignments between the principal axes of the large-scale tidal fields and the spin axes of those present 
galaxies which have preserved well the initially induced tidal influences. From here on, we use ${\bf T}$ to denote 
interchangeably both of the linear tidal field in the proto-galactic stage and the present large-scale tidal field, and  ${\bf j}$ to 
denote the spin axes of the present galaxies without confusing the readers. 

Rearranging the terms of Equation (\ref{eqn:cov}), one can derive the following relation between the tidal link parameter 
$c_{t}$ and the eigenvalues of $\hat{\bf T}$ \citep{LP00,LP01}: 
\begin{equation}
\label{eqn:ct}
c_{t} = \frac{10}{3} - 3\sum_{i=1}^{3}{\vert\hat{\lambda}_{i}\vert^{2}\vert\hat{\bf p}_{i}\cdot{\bf j}\vert^{2}}\, ,
\end{equation}
which implies that $c_{t}$ would vary from realizations to realizations (or from galaxies to galaxies),  
since each realization would have a different degree of the misalignment between the principal axes of ${\bf T}$ 
and ${\bf I}$. 

Suppose that one has a sample of $n_{r}$ galaxies with each having uniquely defined spin vector and that the 
eigenvalues and eigenvectors of the large-scale tidal tensor at the position of each galaxy are all specified. 
Applying Equation (\ref{eqn:ct}) to this sample will give $n_{r}$ different values of $c_{t}$.  
The strength of the mean correlation between the spin vectors of the $n_{r}$ galaxies and the eigenvectors of their 
surrounding tidal field will be determined by the ensemble average, $\langle c_{t}\rangle$, and its error, 
$\sigma_{c_{t}} = \sqrt{20/[9(n_{r}-1)]}$ \citep{LP01}.
Provided that the values of $\bar{c}_{t}$, and $\{\hat{\lambda}_{i}\}_{i=1}^{3}$ are all specified, 
the LTT model will permit a fully analytical evaluation of the three probability density functions, 
$p(\vert\hat{\bf p}_{1}\cdot{\bf j}\vert),\ p(\vert\hat{\bf p}_{2}\cdot{\bf j}\vert)$ and 
$p(\vert\hat{\bf p}_{3}\cdot{\bf j}\vert)$ through Equations (\ref{eqn:pro_j})-(\ref{eqn:pro_hjp}). 

\subsection{The galaxy spin alignments in the cosmic sheets}

The tidal link parameter, $c_{t}$, depends not only on the position of a galaxy but also on the smoothing scale of 
${\bf T}$. For a galaxy with mass $M_{g}$, the parameter $c_{t}$ attains its maximum possible value when ${\bf T}$  
is smoothed on the same scale $M_{g}$, and decreases from the maximum value as the smoothing scale departs 
from $M_{g}$. However, if the tidal field possesses a large-scale coherence in the orientations of their principal axes, 
$c_{t}$ would not substantially fluctuate with the variation of the smoothing scales.  Even when ${\bf T}$ is smoothed 
on a much larger mass scale than $M_{g}$, the correlation strength between ${\bf j}$ and ${\bf T}$ would not sharply 
decrease. 

The coherence of ${\bf T}$ often leads to a formation of the large-scale structure whose constituent galaxies 
exhibit highly anisotropic spatial distributions. Both of a sheet and a filament are the most representative large-scale 
structures that reflects the coherence in the principal directions of ${\bf T}$. Between the two, however, it is only the sheet 
that still remains in the quasi-linear regime \citep{che-etal16}. 
The simplest theory with highest efficacy to describe the formation of a cosmic sheet is the Zel'dovich approximation, 
according to which the ratio of the present to the initial matter density, $\rho/\bar{\rho}$, at a given region is expressed in 
terms of three eigenvalues of ${\bf T}$ as \citep{zel70}: 
\begin{equation}
\label{eqn:zel}
\frac{\rho}{\bar{\rho}} = \frac{1}{(1-\lambda_{1})(1-\lambda_{2})(1-\lambda_{3})}\, .
\end{equation}

Equation (\ref{eqn:zel}) carries two crucial implications: First, a cosmic sheet forms through the compression of 
matter along $\hat{\bf p}_{1}$ at a region where $\lambda_{1}$ comes close to unity when ${\bf T}$ is smoothed on the 
sheet scale.  The fate of a sheet depends on the signs of the other two eigenvalues.  If $\lambda_{2}$ and $\lambda_{3}$ are 
negative when $\lambda_{1}\sim 1$, then the matter in the sheet would never compress along the other two principal axes, 
retaining its two dimensional sheet-like shape for ever. The great wall and the sloan wall must fall into this category 
\citep{great_wall, sloan_wall}. 
In contrast, for the case of $\lambda_{2}\ge \lambda_{3}\ge 0$, then the region will first evolve into a sheet via the 
one dimensional compression along $\hat{\bf p}_{1}$. However, in the subsequent process, it will also collapse along 
$\hat{\bf p}_{2}$ and $\hat{\bf p}_{3}$ to evolve into a filament and eventually into a knot, respectively. 
\citet{lee04} defined a cosmic sheet of the second type as a structure that forms at a region where the smallest 
tidal eigenvalue, $\lambda_{3}$, just reaches zero. To justify this definition, \citet{lee04} computed the 
expectation value of $\lambda_{1}$ under the condition of $\lambda_{3}=0$ and found 
$\langle \lambda_{1}\vert \lambda_{3}\rangle\sim 0.9$ (i.e., the condition for the one dimensional collapse along 
$\hat{\bf p}_{1}$ according to the Zel'dovich approximation), regardless of the smoothing scale. 

It is worth mentioning why the definition of a sheet suggested by \citet{lee04} and adopted here seems apparently different 
from the the conventional one according to which a sheet corresponds to a region where the local tidal tensor 
smoothed on a certain scale has only one positive eigenvalue \citep[e.g.,][]{hah-etal07,zha-etal15}. The difference lies in the 
fact that the eigenvalues in the two definitions are not the same ones. In the latter the eigenvalues are the Eulerian quantities 
describing the matter distribution surrounding a sheet after its formation via the one dimensional compression, 
while in the latter the eigenvalues, i.e., $\{\lambda_{i}\}_{i=1}^{3}$, that appear in Equations (\ref{eqn:pro_hj})-(\ref{eqn:zel}) 
are not the Eulerian but the Lagrangian quantities describing the surrounding matter distribution before the formation of a sheet.  
 
Combining the condition of $\lambda_{3}=0$ with the other two constraints of 
$\sum{\hat{\lambda}^{2}_{i}}=1$ and $\sum{\hat{\lambda}_{i}}=0$, one can show that 
$\{\hat{\lambda}_{i}\}^{3}_{i=1}$ can be written in terms only of $\lambda_{1}$ and $\lambda_{2}$ as \citep{lee04}: 
\begin{eqnarray}
\label{eqn:hl1}
\hat{\lambda}_{1} &=& \frac{2\lambda_{1}-\lambda_{2}}
{\sqrt{6(\lambda^{2}_{1}-\lambda_{1}\lambda_{2}+\lambda^{2}_{2}})}\, , \\
\label{eqn:hl2}
\hat{\lambda}_{2} &=& 
{\rm max}\bigg{\{}\frac{2\lambda_{2}-\lambda_{1}}
{\sqrt{6(\lambda^{2}_{1}-\lambda_{1}\lambda_{2}+\lambda^{2}_{2}})}, \  
\frac{-(\lambda_{1}+\lambda_{2})}{\sqrt{6(\lambda^{2}_{1}-\lambda_{1}\lambda_{2}+\lambda^{2}_{2}})}\bigg{\}}\, , \\
\label{eqn:hl3}
\hat{\lambda}_{3} &=& 
{\rm min}\bigg{\{}\frac{2\lambda_{2}-\lambda_{1}}
{\sqrt{6(\lambda^{2}_{1}-\lambda_{1}\lambda_{2}+\lambda^{2}_{2}})}, \  
\frac{-(\lambda_{1}+\lambda_{2})}{\sqrt{6(\lambda^{2}_{1}-\lambda_{1}\lambda_{2}+\lambda^{2}_{2}})}\bigg{\}}\, .
\end{eqnarray}

The second implication of Equation (\ref{eqn:zel}) is that the observed shape of a cosmic sheet can be used to determine 
the eigenvalues of the surrounding large-scale tidal field.  Suppose that a sheet has a shape of a rectangular parallelepiped 
with side lengths $L_{1},\ L_{2},\ L_{3}$ with $L_{1}\ge L_{2}\ge L_{3}$.  The longest, second longest and shortest sides of this 
parallelepiped-like sheet must be in parallel to the minor ($\hat{\bf p}_{3}$), intermediate ($\hat{\bf p}_{2}$) and 
major ($\hat{\bf p}_{1}$) principal axes of ${\bf T}$, respectively, with its present volume given as 
$V_{\rm sheet}=\prod_{i=1}^{3}{L_{i}}$. 
Let $V_{L}$ denote the volume of a cosmic sheet in its initial stage before the completion of the gravitational collapse 
along $\hat{\bf p}_{1}$ where the matter density of the sheet equals the mean background density. 
Since the mass contained in a volume is always conserved, one can say 
$\rho V_{\rm sheet} = \rho\prod_{i=1}^{3}{L_{i}}=\bar{\rho} V_{L}$.  By Equation (\ref{eqn:zel}), 
we now have  the following relation between $\{L_{i}\}_{i=1}^{3}$ and $\{\lambda_{i}\}_{i=1}^{3}$
\begin{equation}
\label{eqn:side}
L_{1}=V_{L}^{1/3}(1-\lambda_{3})\, , \quad
L_{2}=V_{L}^{1/3}(1-\lambda_{2})\, , \quad
L_{3}=V_{L}^{1/3}(1-\lambda_{1})\, .
\end{equation}

Putting $\lambda_{3}=0$ (the sheet criterion) into Equation (\ref{eqn:side}), one can determine the 
value of the proportionality constant, $V_{L}^{1/3}$. Putting the determined value of $V_{L}^{1/3}$ back into Equation 
(\ref{eqn:side}),  one can obtain the values of $\lambda_{2}$ and $\lambda_{1}$, which can in turn be used to 
find the values of $\{\hat{\lambda}_{i}\}^{3}_{i=1}$ through Equations (\ref{eqn:hl1})-(\ref{eqn:hl3}) and the value of 
$c_{t}$ by Equation (\ref{eqn:ct}).  
Finally, a complete analytic evaluation of the prediction of the LTT model on the intrinsic spin alignments of the 
galaxies embedded in a cosmic sheet can be made by putting $\{\hat{\lambda}_{i}\}^{3}_{i=1}$ and 
$\bar{c}_{t}$ into Equation (\ref{eqn:pro_hjp}), where $\bar{c}_{t}$ is the average of $c_{t}$ taken over the constituent 
galaxies of the sheet.

\section{Testing the LTT model with the WM Sheet}\label{sec:test}

The WM sheet is a sheet-like structure composed of $204$ galaxies (called the WMs galaxies hereafter
\footnote{\citet{kim-etal16} found that the WM sheet comprise a total of $220$ galaxies. However, information on the 
position angle and axial ratios are available only for $204$ WMs galaxies.}) discovered by 
\citet{kim-etal16} in the vicinity of the Virgo cluster and the Local Void \citep[see Figures 1 and 2 in][]{kim-etal16}. 
The ranges of the supergalactic Cartesian coordinates, ($x_{sg},\ y_{sg},\ z_{sg}$) of the WMs galaxies are listed in Table 
\ref{tab:wm}. Although it is located in the neighborhood of the Virgo cluster, a careful analysis of the recession velocities of 
the WMs galaxies has revealed that the WM sheet does not gravitationally interact with the Virgo cluster
\citep{kim-etal16}.  

The top-left, top-right and bottom-left panels of Figure \ref{fig:wm_sg} show the projected positions of the WMs 
galaxies onto the supergalactic $\hat{\bf x}$-$\hat{\bf y}$, $\hat{\bf y}$-$\hat{\bf z}$ and $\hat{\bf x}$-$\hat{\bf z}$ 
planes, respectively.  
As can be seen, the WM sheet has an interesting property: its shape is well approximated as a rectangular 
parallelepiped with its longest, second longest, and shortest sides well aligned the supergalactic $\hat{\bf x}$, 
$\hat{\bf y}$ and $\hat{\bf z}$ axes, respectively.  Given this observed property of the WM sheet the principal axes 
of the surrounding large-scale tidal field seem to coincide with the supergalactic Cartesian axes as: 
$\hat{\bf p}_{1}\parallel \hat{\bf z},\ \hat{\bf p}_{2}\parallel \hat{\bf y},\ \hat{\bf p}_{3}\parallel \hat{\bf x}$, 
since the shortest and longest sides of a sheet are expected to be aligned with the directions of the strongest 
and the weakest tidal forces.

Extracting information on the position angles ($\psi$) and axial ratios ($R_{\rm axial}$) of the WMs galaxies obtained 
through the sersic fitting from the website of the NASA-Sloan-Atlas galaxy catalog 
\footnote{http://www.nsatlas.org/data}, we adopt the conventional methodology to determine the unit spin vector 
of each WMs galaxy in the equatorial Cartesian coordinate system as 
\citep{LP00,pen-etal00,tru-etal06,lee11,var-etal12,tem-etal13}
\begin{equation}
\label{eqn:jx}
\begin{pmatrix}
j_{x}\\
j_{y}\\
j_{z}
\end{pmatrix}_{\rm equatorial}
=
\begin{pmatrix}
j_{r}\sin\theta\cos\phi + j_{\theta}\cos\theta\cos\phi  - 
j_{\phi}\sin\phi \\
j_{r}\sin\theta\sin\phi  + j_{\theta}\cos\theta\sin\phi + 
j_{\phi}\cos\phi \\
j_{r}\cos\theta - j_{\theta}\sin\theta 
\end{pmatrix}\, ,
\end{equation}
where the spherical polar components of the unit spin vector, $j_{r},\ j_{\theta},\ j_{\phi}$, are 
given in terms of $R_{\rm axis}$ and $\psi$ as
\begin{equation}
\label{eqn:jr}
j_{r}=\pm R_{\rm axial}\, , \quad j_{\theta}=\sqrt{1-\cos^{2}\xi}\sin\psi\, , \quad
j_{\phi}=\sqrt{1-\cos^{2}\xi}\cos\psi\, .
\end{equation}
The plus and minus signs before $R_{\rm axial}$ indicate our inability to determine whether the rotation of a 
WMs galaxy upon its spin axis is clock or counterclock wise. This two-fold sign ambiguity in the determination of 
${\bf j}$ increases the total number of the realizations of ${\bf j}$ to twice the total number of the WMs 
galaxies. That is, we have a total of $408$ realizations of ${\bf j}$ from the WM sheet. 

The unit spin vector ${\bf j}$ in the supergalactic Cartesian coordinate system can be obtained through 
the rotation of the coordinate system as
\begin{equation}
\begin{pmatrix}
j_{x}\\
j_{y}\\
j_{z}
\end{pmatrix}_{\rm supergalactic}
=
\begin{pmatrix}
R_{11} & R_{12} & R_{13} \\
R_{21} & R_{22} & R_{23} \\
R_{31} & R_{32} & R_{33} 
\end{pmatrix}^{-1}
\begin{pmatrix}
j_{x}\\
j_{y}\\
j_{z}
\end{pmatrix}_{\rm equatorial}
\end{equation}
where ${\bf R}^{-1}$ is the inverse (or transponse) of the rotation matrix ${\bf R}=(R_{ij})$ whose nine components can 
be obtained from the J2000 values of the right ascension and declination of the supergalactic North pole  
($\alpha_{n}=2.82\,$h, $\delta_{n}=+59.5^{\circ}$) and those of the supergalactic zero point line 
($\alpha_{0}=2.82\,$h, $\delta_{0}=+59.5^{\circ}$): 
\begin{equation}
\left(R_{ij}\right) = 
\begin{pmatrix} 
\cos\delta_{0}\cos\alpha_{0} &  \sin\delta_{n}\cos\alpha_{n}\sin\alpha_{0} - \sin\alpha_{n}\sin\delta_{0}\cos\alpha_{0} & 
\cos\delta_{n}\cos\alpha_{n} \\
\sin\delta_{0}\cos\alpha_{0} & 
\sin\alpha_{n}\cos\delta_{0}\cos\alpha_{0}-\cos\delta_{n}\cos\alpha_{n}\sin\alpha_{0} &
\sin\delta_{n}\cos\alpha_{n} \\
\sin\alpha_{0} &
\cos\alpha_{n}\cos\alpha_{0}\sin(\delta_{0}-\delta_{n}) &
\sin\alpha_{n}
\end{pmatrix}\, .
\end{equation}
From here on, ${\bf j}=(j_{x},\ j_{y},\ j_{z})$ denote exclusively the spin vector of a WMs 
galaxy measured in the supergalactic Cartesian coordinate system. 

The absolute value of the $\hat{\bf x}$-axis component of ${\bf j}$, $\vert j_{x}\vert$, is in the range of 
$[0, 1]$. Breaking this unit interval into four bins of equal size $1/4$, we count the number of the realizations 
falling in each bin to determine in a discrete way the probability density function of $\vert j_{x}\vert$ as 
$p(\vert j_{x}\vert_{i})=4n_{i}/N_{\rm r}$ where $\vert j_{x}\vert_{i}$ and $n_{i}$ are the value of 
$\vert j_{x}\vert$ and the number of the realizations at the $i$-th bin, respectively, and $N_{\rm r}$ is 
the total number of the realizations.  The errors are estimated by employing the Bootstrap method as 
\begin{equation}
\label{eqn:boot}
\sigma^{2}_{x,i} = \frac{1}{N_{\rm boot}}
\sum_{k=1}^{N_{\rm boot}}{\left[p^{k}(\vert j_{x}\vert_{i})-p(\vert j_{x}\vert_{i})\right]^{2}}\, ,
\end{equation}
where $\sigma_{x,i}$ is the error in the measurement of $p(\vert j_{x}\vert_{i})$ at the $i$-th bin, 
$N_{\rm boot}=10000$ is the number of the Bootstrap resamples, and $p^{k}(\vert j_{x}\vert_{i})$ is the 
probability density at the $i$-th bin from the $k$-th Bootstrap resample. 
The other two probability density functions, $p(\vert j_{y}\vert)$ and $p(\vert j_{z}\vert)$, along with 
their errors are all determined in the same manner. 

The top-left, top-middle and top-right panels of Figure \ref{fig:pro_all} show this observationally determined probability 
density functions, $p(\vert j_{x}\vert)$, $p(\vert j_{y}\vert)$ and $p(\vert j_{z}\vert)$, respectively, as filled black circles with the 
Bootstrap errors. 
Since $\hat{\bf x}\parallel \hat{\bf p}_{3}$, $\hat{\bf y}\parallel \hat{\bf p}_{2}$, $\hat{\bf z}\parallel \hat{\bf p}_{1}$, 
the observational results shown in Figure \ref{fig:pro_all} are in fact the probability density functions of the absolute 
values of the components of the unit spin vectors of the WMs galaxies along the minor, intermediate and major 
principal axes of the surrounding large-scale tidal field, respectively. That is, 
$p(\vert j_{x}\vert)=p(\vert\hat{\bf p}_{3}\cdot{\bf j}\vert)$, 
$p(\vert j_{y}\vert)=p(\vert\hat{\bf p}_{2}\cdot{\bf j}\vert)$ and 
$p(\vert j_{z}\vert)=p(\vert\hat{\bf p}_{1}\cdot{\bf j}\vert)$. 

As can be seen in Figure \ref{fig:pro_all}, $p(\vert j_{x}\vert)$ appears to be more or less uniform.  
No consistent trend of increment nor decrement with $\vert j_{z}\vert$ is found in the shape of 
$p(\vert j_{z}\vert)$, either. But, its abrupt deep drop at the fourth 
bin from the right implies that it is less probable to find the WMs galaxies whose spin vectors are aligned with 
the major principal axis of the surrounding tidal field. Meanwhile, a clear signal of the preferential alignment of 
${\bf j}$ with $\hat{\bf p}_{2}$ is found in the shape of $p(\vert j_{y}\vert)$, 
which seems qualitatively consistent with the prediction of the LTT model.

The bottom panels of Figure \ref{fig:pro_all} show the corresponding number distributions of the realizations of the spin axes 
of the WM galaxies as histograms with the Poisson errors. Strictly speaking, both of the bootstrap and Poisson errors are the 
underestimation of the true uncertainties associated with the measurements of the probability density or equivalently the number  
distributions. 
It is because the two realizations of the spin axis of each WM galaxy are not independent with each other but differs from each 
other only in the sign of $j_{r}$, while the bootstrap and Poisson errors have been estimated under the assumption that the 
two realizations are mutually independent.  However, it does not necessarily mean that the true uncertainties 
should be larger than the Poisson errors by a factor of $\sqrt{2}$. Although the two realizations are correlated with each other, 
their orientations with respect to the principal directions of the linear tidal field could be quite different from each other. 
The true spin axis of each WM galaxy is either one of the two realizations. If one realization is the real spin axis being aligned with 
the principal direction of the linear tidal field, then the other realization that has an opposite sign in $j_{r}$ is likely to have no 
tidal alignment tendency. This is why the previous works that statistically investigated the tidal alignments of the galaxy spin axes 
\citep[e.g.,][]{pen-etal00,var-etal12,TL13} adopted the assumption that the two realizations are mutually independent. 
As a matter of fact, \citet{var-etal12} justified this assumption with the help of a Monte-carlo simulation. 

Now that $p(\vert j_{x}\vert),\ p(\vert j_{y}\vert),\ p(\vert j_{z}\vert)$ are all observationally determined, 
we would like to make a quantitative comparison of them with the prediction of the LTT model. 
Taking the differences between the maximum and minimum values of the supergalactic Cartesian coordinates of 
each WMs galaxies (see Table \ref{tab:wm}), we measure the side lengths of the WM sheet as
\begin{equation}
\label{eqn:lrange}
L_{3} = (x_{\rm sg,max} - x_{\rm sg,min})\, ,\quad
L_{2} = (y_{\rm sg,max} - y_{\rm sg,min})\, ,\quad
L_{1} = (z_{\rm sg,max} - z_{\rm sg,min})\, ,
\end{equation}
which gives $L_{3}=11.22$\,Mpc, $L_{2}=8.95$\,Mpc, and $L_{1}=1.32$\,Mpc.  Plugging these values as well as 
the sheet-condition of $\lambda_{3}=0$ into Equation (\ref{eqn:side}), we find
$\lambda_{1}=0.88$, $\lambda_{2}=0.2$ and $V^{1/3}_{L}=11.21$Mpc. 
Putting the values of $\lambda_{1}$ and $\lambda_{2}$ into Equations (\ref{eqn:hl1})-(\ref{eqn:hl3}), 
we finally draw $\hat{\lambda}_{1}=0.797,\ \hat{\lambda}_{2}=-0.243$ and $\hat{\lambda}_{3}=-0.553$. 

Plugging the measured values of $\{\hat{\lambda}_{i}\}_{i=1}^{3}$ along with the three components of the unit 
spin vectors into Equation (\ref{eqn:ct}), we finally determine the $408$ different values of $c_{t}$ for the same 
number of the ${\bf j}$-realizations. 
The top left panel of Figure \ref{fig:ct} displays the probability density function of $c_{t}$, which seems to reach its 
maximum around $c_{t}=0.1$, exhibiting a widely spread asymmetric shape over the range of $[-3,3]$. 
The first row of Table \ref{tab:ct} lists the mean value, $\bar{c}_{t}$,  averaged over the $408$ realizations and its 
associated error $\sigma_{c_{t}}$, revealing a detection of a statistically significant signal of the correlation between 
${\bf j}$ and $\hat{\bf T}$ from the WM sheet. 

Putting the measured values of $\{\hat{\lambda}_{i}\}_{i=1}^{3}$ and $\bar{c}_{t}$ into 
Equations (\ref{eqn:pro_hj})-(\ref{eqn:pro_hjp}), we analytically evaluate the probability density functions of the 
components of the unit spin vectors of the WMs galaxies along the major, intermediate and minor principal axes of the 
surrounding large-scale tidal field. which are displayed as green area in Figure \ref{fig:pro_all}.  
Recalling that $\vert j_{x}\vert=\vert\hat{\bf p}_{3}\cdot{\bf j}\vert,\
\vert j_{y}\vert=\vert\hat{\bf p}_{2}\cdot{\bf j}\vert,\ 
\vert j_{z}\vert=\vert\hat{\bf p}_{1}\cdot{\bf j}\vert$, we note that the observational results (filled circular dots) do not agree  
with the analytic prediction (green area) at the quantitatively level.

We suspect that this failure of the LTT model in quantitively matching the observational results should be caused 
by the development of the non-linearity in the dense inner region of the WM sheet.  The LTT model which assumes the 
Gaussianity of $p({\bf j})$ is strictly valid only in the linear regime where the matter density rms fluctuation does not 
exceeds unity.  Equation (\ref{eqn:cov}) which measures the covariance among the components of the unit spin 
vectors of the galaxies is also strictly effective only in the linear regime \citep{LP08}. Furthermore, the Zel'dovich 
approximation from which the sheet-criterion of $\lambda_{3}=0$ is extracted also breaks down in the nonlinear 
regime \citep{sha-etal94}. 

In the top-left panel of Figure \ref{fig:wm_sg}, one can glimpse of the presence of a filament-like substructure aligned 
with the supergalactic $\hat{\bf y}$-axis in the inner section of the WM sheet, which actually indicates that the inner 
overdense section of the WM sheet may have already begun gravitational collapse along $\hat{\bf p}_{2}$, 
entering a nonlinear regime. To eliminate the contamination caused by the existence of this nonlinear substructure, 
we first find the center of the WMs galaxies, say ${\bf x}_{sg,c},\  {\bf y}_{sg,c},\ {\bf z}_{sg,c}$. 
Then, we measure the distance vector, ${\bf d} = (d_{x},d_{y},d_{x})$, of each WMs galaxy from the center 
as $d_{x}= \vert x_{sg}-x_{sg,c}\vert$, $d_{y}=\vert y_{sg}-y_{sg,c}\vert$, $d_{z}=\vert z_{sg}-z_{sg,c}\vert$.  
Then, we repeat the whole analysis with only those WMs galaxies that satisfy $d_{x} \ge d_{x,c}$ where 
$d_{x,c}\sim 1.2$ Mpc is the median value of the $\hat{\bf x}$-axis components of the distance vectors of the WMs galaxies. 
The left panel of Figure \ref{fig:wm_core} shows the configurations of the WM galaxies satisfying the condition of 
$d_{x} \ge d_{x,c}$ as filled red dots in the supergalactic plane. 

The top-right panel of Figure \ref{fig:ct} plots the same as the top-left but for the case of only those WMs galaxies 
with $d_{x}\ge d_{x,c}$.  The $c_{t}$-distribution exhibits a more asymmetric shape, being biased toward 
the positive values of $c_{t}$.  Figure \ref{fig:pro_dxh} displays the same as Figure \ref{fig:pro_all} but for this 
case of  $d_{x}\ge d_{x,c}$.  As can be seen, although the mean value of $c_{t}$ for this case 
(listed in the second raw of Table \ref{tab:ct}) is found to decrease, the three probability density functions, 
$p(\vert j_{x}\vert),\ p(\vert j_{y}\vert),\ p(\vert j_{z}\vert)$, show much better agreements with the analytic 
prediction of the LTT theory. 

Repeating the whole calculations with those WMs galaxies that satisfy $d_{x}<d_{x,c}$, we show the corresponding 
results in  the bottom left panel of Figure \ref{fig:ct}, Figure \ref{fig:pro_dxl} and third raw of Table \ref{tab:ct}.
As can be seen,  although the alignment tendency between ${\bf j}$ and $\hat{\bf p}_{2}$ becomes stronger, 
the overall agreements between the observational and the analytical probability density functions degrade significantly. 
Especially, the weak ${\bf j}$-$\hat{\bf p}_{2}$ alignment and the strong ${\bf j}$-$\hat{\bf p}_{1}$ anti-alignments 
predicted by the LTT model seem to disappear in this observational result for the case of $d_{x}<d_{x,c}$.   

Noting that the exclusion of the inner dense section from the WM sheet is important for the detection of a true 
linear tidal vestige, we measure the distance of each WM galaxy from the center as 
$d=\sqrt{d^{2}_{x}+d^{2}_{y}+d^{3}_{z}}$ and redo the analysis with those WMs galaxies satisfying 
$d\ge d_{c}\sim 2.2\,$Mpc, where $d_{c}$ represents a typical nonlinear scale. 
The right panel of Figure \ref{fig:wm_core} shows the 2D configuration of these WM galaxies with $d\ge d_{c}$ 
in the supergalactic plane as filled red dots. 
The bottom-right panel of Figure \ref{fig:ct},  Figure \ref{fig:pro_d} and the fourth raw of Table \ref{tab:ct} show the 
corresponding results of this reanalysis. As can be seen, the exclusion of $d< d_{c}$ yields even a better quantitative 
agreement between the observational and the analytic restuls than that for the case of $d_{x}\ge d_{x,c}$. 
The mean parameter, $\bar{c}_{t}$, turns out to increase up to $0.29\pm 0.01$, confirming that the linear tidal 
vestige remains in the WM sheet at a statistically significant level. 

It is worth discussing why the WM galaxies with $d\ge d_{c}$ yield a stronger signal of the linear tidal alignment 
than those with $d_{x}\ge d_{x,c}$, even though the substantial fraction of the WM galaxies with $d\ge d_{c}$ 
obviously belong to the filamentary section defined as $d_{x}<d_{x,c}$ (see Figure \ref{fig:wm_core}). 
Although the WM sheet as a whole is a quasi-linear structure, its densest core enclosed by $d_{c}$ is believed to have 
undergone a severe nonlinear evolution. In consequence the spin directions of the WM galaxies located in the central core 
section with $d<d_{c}$ must have almost completely lost their initial tendency of being aligned with the principal directions of 
the linear tidal field. 
The section with $d\ge d_{c}$ does not contain those WM galaxies whose spin directions have lost their memory 
of the linear tidal effect, while the section with $d_{x}\ge d_{x,c}$ contains such galaxies, which explains why 
the former yields a better agreement with the analytic prediction of the LTT model than the latter. 
Our results shown in Figures \ref{fig:pro_dxh}-\ref{fig:pro_d} imply that the nonlinear effect on the spin directions of the WM 
galaxies in the filament section with $d_{x}<d_{x,c}$ would not be so strong as those in the central core section with 
$d\ge d_{c}$.

\section{Summary and Discussion}\label{sec:sum}

\begin{figure}
\begin{center}
\epsscale{1.0}
\plotone{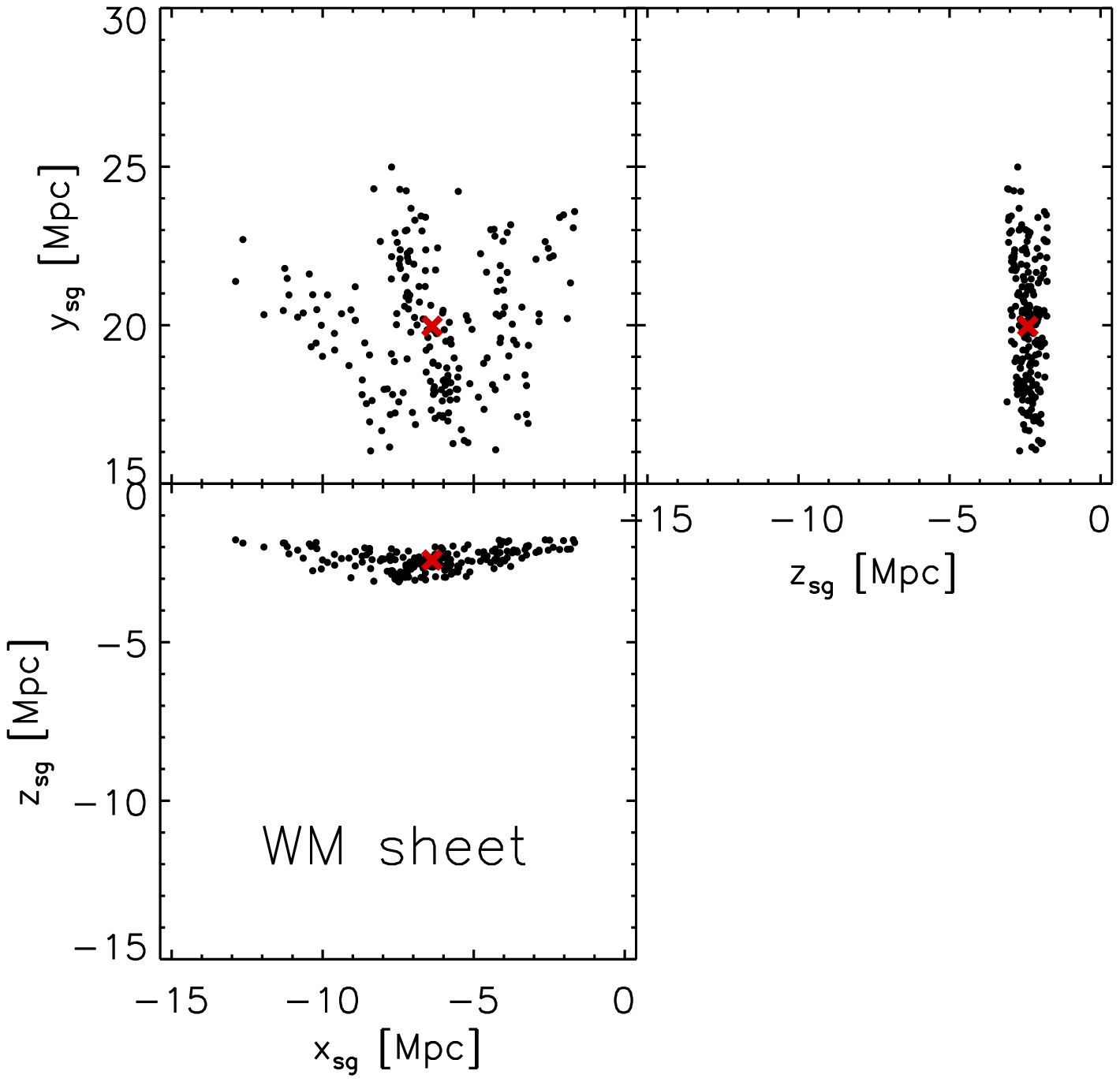}
\caption{Configuration of the constituent galaxies of the WM sheet in the supergalactic coordinate system. The red cross 
in each panel represents the projected position of the center of the WM sheet.}
\label{fig:wm_sg}
\end{center}
\end{figure}

\begin{figure}
\begin{center}
\epsscale{1.0}
\plotone{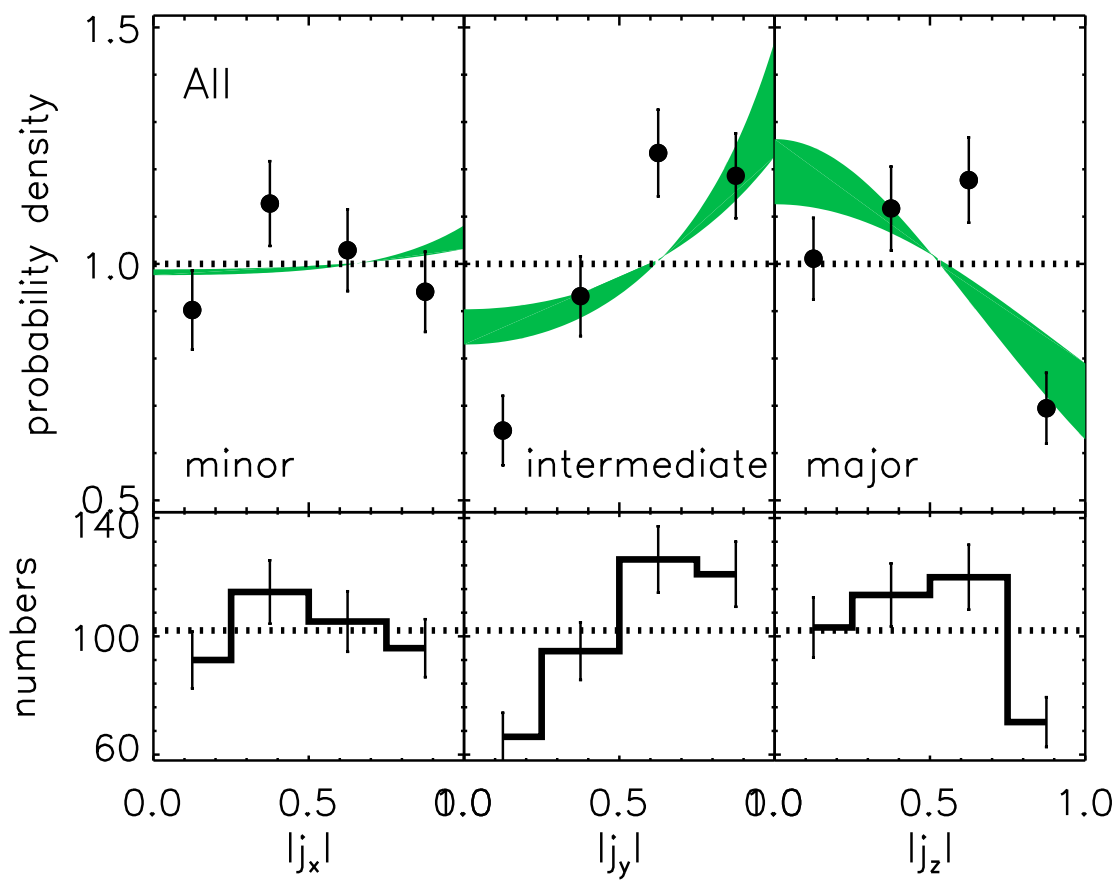}
\caption{Probability density functions of the ${\bf x}$, ${\bf y}$ and ${\bf z}$-axis components of the unit spin vectors, 
${\bf j}$, of all WMs galaxies are displayed as filled circular dots with the bootstrap errors in the top-left, top-middle and 
top-right panels, respectively. In each top panel, the green area corresponds to the theoretical prediction of the LTT model. 
The corresponding distributions of the realization numbers are also shown as histograms with the Poisson errors in the bottom 
panels. }
\label{fig:pro_all}
\end{center}
\end{figure}

\begin{figure}
\begin{center}
\epsscale{1.0}
\plotone{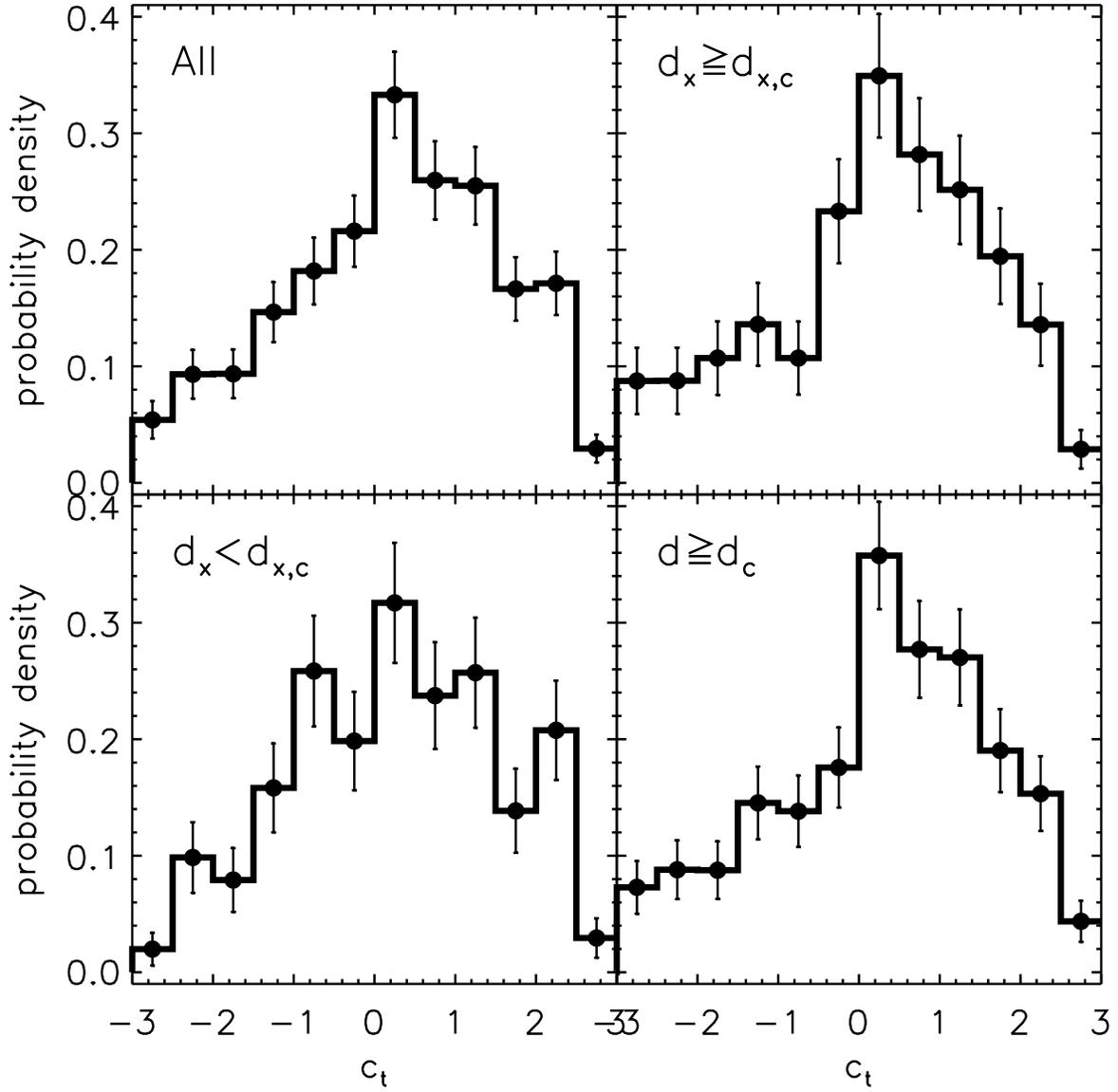}
\caption{Probability density functions of the tidal link parameters $c_{t}$.}
\label{fig:ct}
\end{center}
\end{figure}

\begin{figure}
\begin{center}
\epsscale{1.0}
\plotone{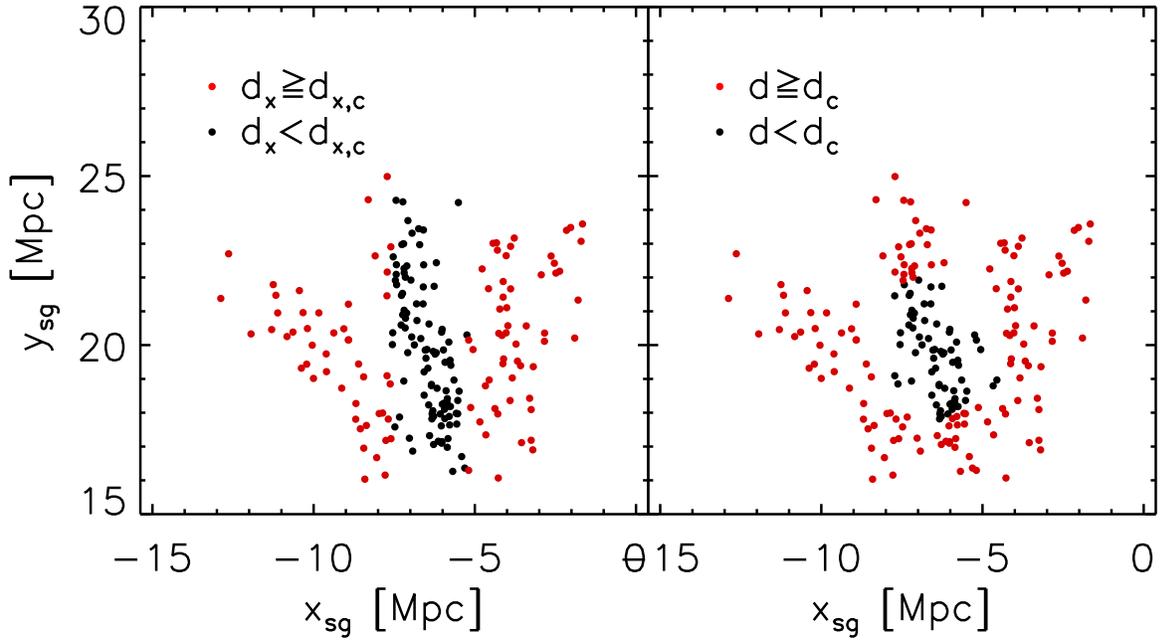}
\caption{Configuration of the WM galaxies with $d_{x}> d_{x,c}$ (left panel) and with $d> d_{c}$ (right panel) 
as filled red dots in the supergalactic plane. In each panel, the configuration of the other WM galaxies are also 
shown as filled black dots.}
\label{fig:wm_core}
\end{center}
\end{figure}

\begin{figure}
\begin{center}
\epsscale{1.0}
\plotone{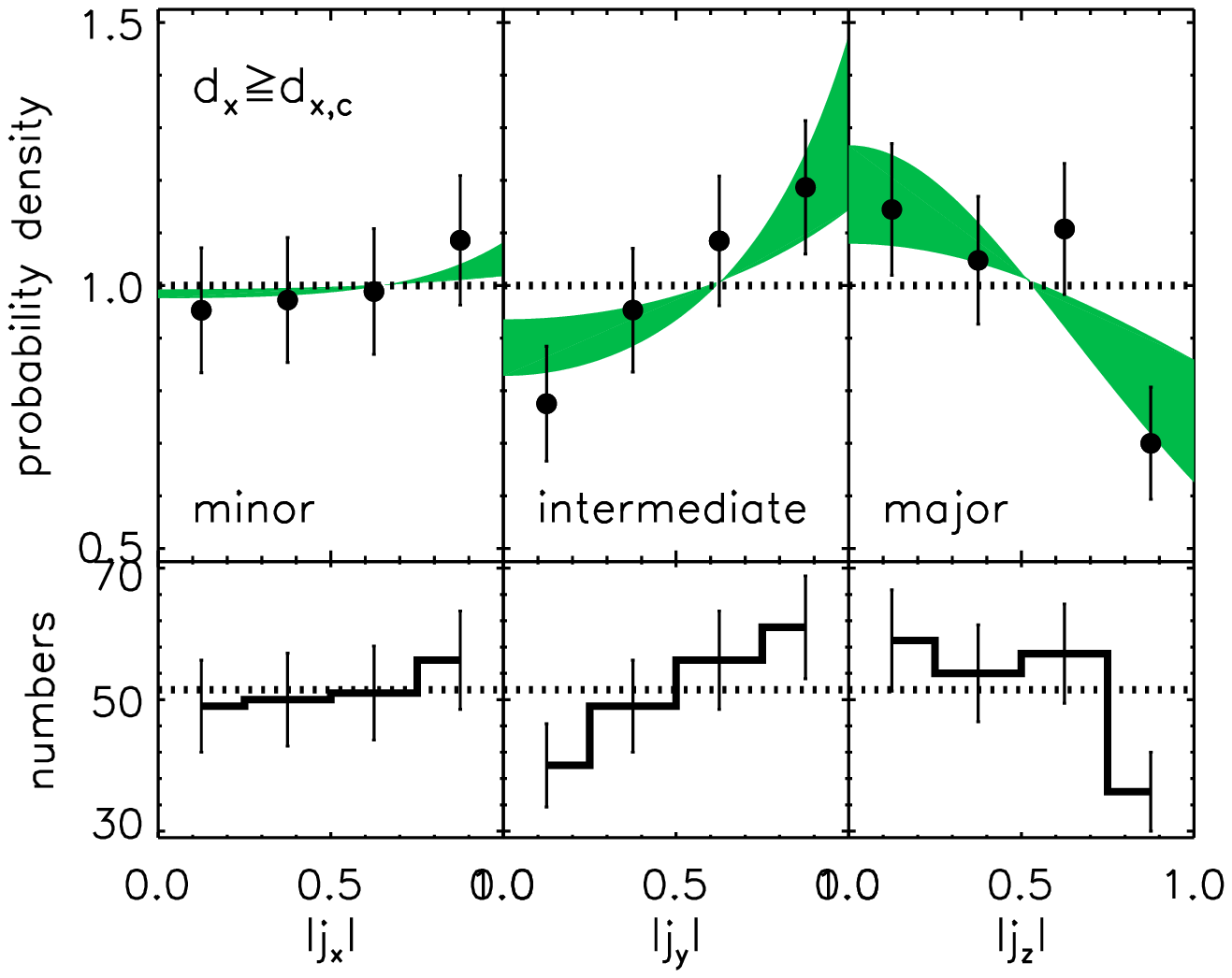}
\caption{Same as Figure \ref{fig:pro_all} but for the case of those WMs galaxies satisfying 
$d_{x}\ge d_{x,c}$.}
\label{fig:pro_dxh}
\end{center}
\end{figure}

\begin{figure}
\begin{center}
\epsscale{1.0}
\plotone{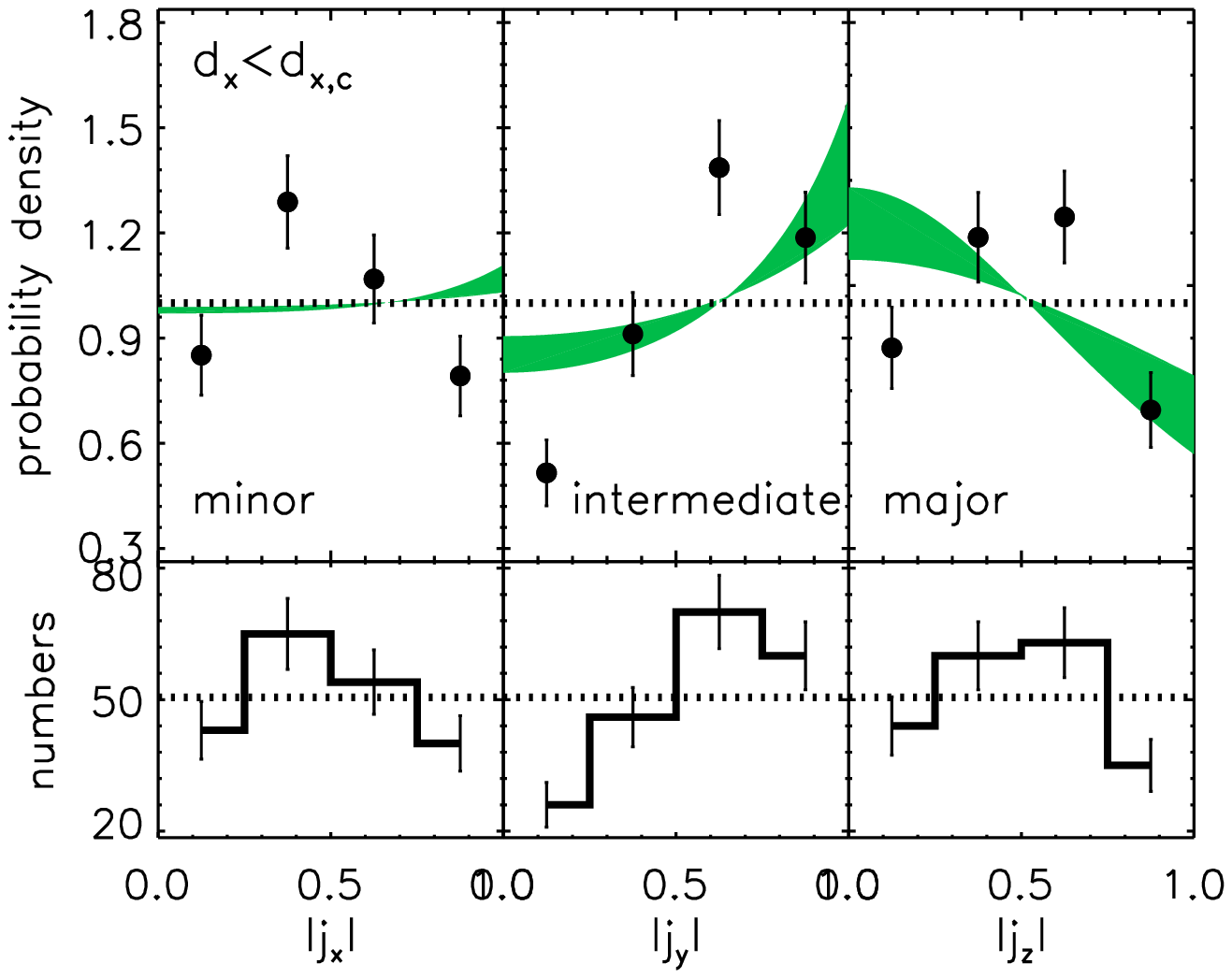}
\caption{Same as Figure \ref{fig:pro_all} but for the case of those WMs galaxies satisfying 
$d_{x}<d_{x,c}$.}
\label{fig:pro_dxl}
\end{center}
\end{figure}

We have tested the LTT (linear tidal torque) model against the WM sheet whose longest, second longest, and shortest 
sides are observed to be parallel with the supergalactic Cartesian $\hat{\bf x},\ \hat{\bf y}$ and $\hat{\bf z}$ axes, respectively 
(Figure \ref{fig:wm_sg}). Measuring the unit spin vectors, ${\bf j}=(j_{x},j_{y},j_{z})$, of the WMs galaxies 
(WM sheet galaxies) in the supergalactic coordinate system according to the conventional methodology, we have 
determined in a discrete way three probability density functions, $p(\vert j_{x}\vert),\ p(\vert j_{y}\vert),\ p(\vert j_{z}\vert)$, 
and found a sharp increase of $p(\vert j_{y}\vert)$ with $\vert j_{y}\vert$, an abrupt deep drop of $p(\vert j_{z}\vert)$ below 
unity at $\vert j_{z}\vert\sim 1$, and no consistent variation of $p(\vert j_{x}\vert)\sim 1$ against $\vert j_{x}\vert$ 
(Figure \ref{fig:pro_all}). 

Noting that the minor ($\hat{\bf p}_{3}$), intermediate ($\hat{\bf p}_{2}$) and major ($\hat{\bf p}_{1}$) principal 
directions of the large-scale tidal field surrounding the WM sheet coincide with the supergalactic 
$\hat{\bf x},\ \hat{\bf y}$ and $\hat{\bf z}$ axes, respectively, we have interpreted this observational result as a 
supporting evidence for the existence of the strong alignment and anti-alignment of ${\bf j}$ with $\hat{\bf p}_{2}$ 
and $\hat{\bf p}_{1}$, which trend is consistent with the prediction of the LTT model \citep{LP00,LP01}, 
under the well justified assumption that the large-scale tidal field surrounding the WM sheet is a good approximation 
to the linear tidal field coherent over the mass scale of the WM sheet. 

Measuring the side lengths of the WM sheet and assuming that the smallest eigenvalue of the linear tidal tensor at the 
center of the WM sheet is zero \citep{lee04}, we have determined the other two (largest and second largest) tidal 
eigenvalues with the help of the Zel'dovich approximation \citep{zel70}. Then, we have measured the mean values of the 
tidal link parameter, $\bar{c}_{t}$, which quantifies the degree of the deviations of the above three probability density 
functions from the uniform distribution.  The obtained values of $\lambda_{1},\ \lambda_{2},\ c_{t}$ have allowed us to 
evaluate $p(\vert j_{x}\vert),\ p(\vert j_{y}\vert),\ p(\vert j_{z}\vert)$ in a full analytic manner according to the LTT model. 

\begin{figure}
\begin{center}
\epsscale{1.0}
\plotone{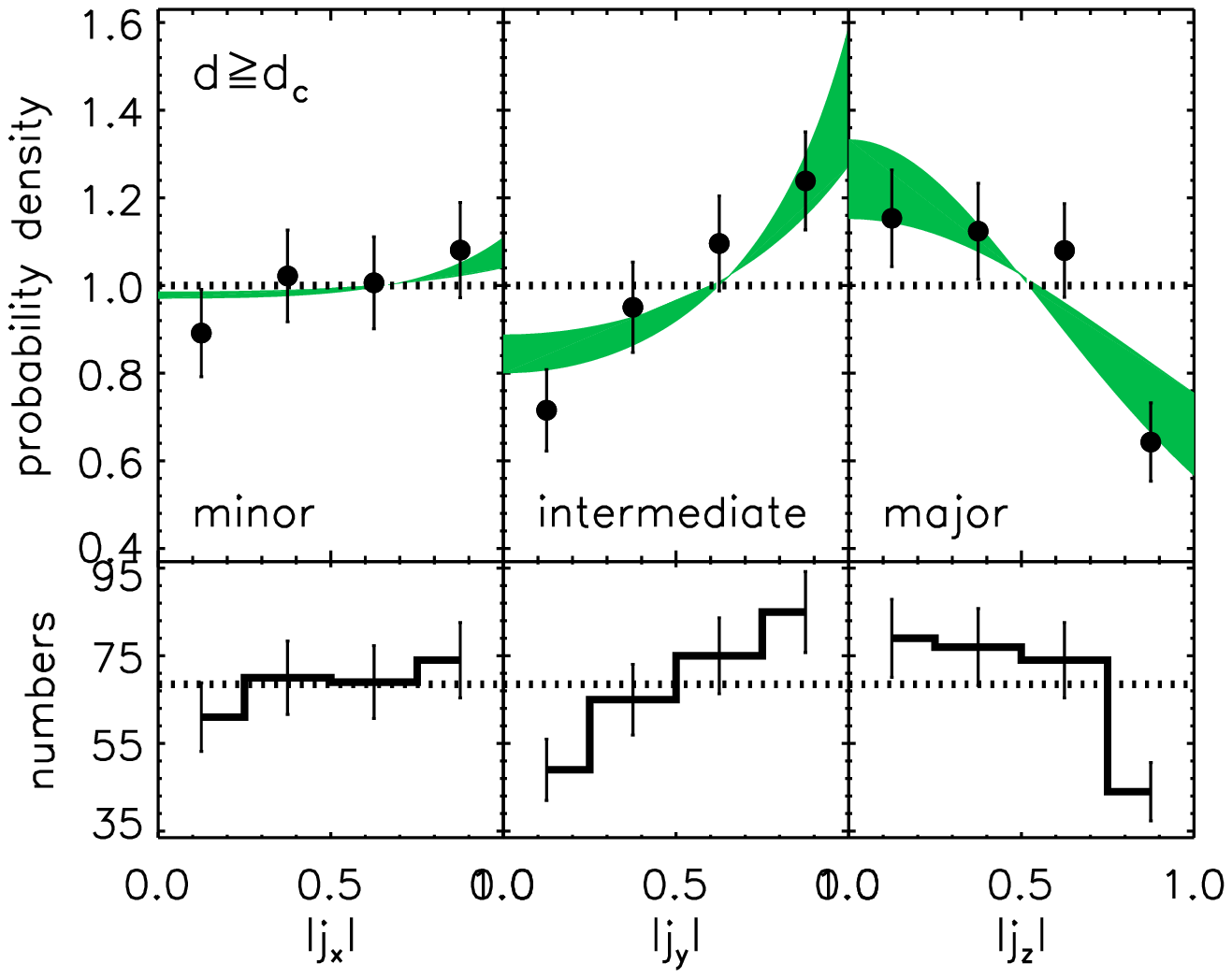}
\caption{Same as Figure \ref{fig:pro_all} but for the case of those WMs galaxies satisfying 
$d\ge d_{c}$.}
\label{fig:pro_d}
\end{center}
\end{figure}

The analytic predictions of the LTT model have failed in matching quantitatively the observed behaviors of the 
three probability density functions. Suspecting that the growth of nonlinearity in the inner denser section of the WM sheet 
should be responsible for this failure of the LTT model, we have excluded those WMs galaxies located in the central core of 
a typical nonlinear scale radius and redone the whole analysis.  All of the three probability density functions, 
$p(\vert j_{x}\vert),\ p(\vert j_{y}\vert),\ p(\vert j_{z}\vert)$, from this reanalysis have been shown to agree quantitatively well 
with the analytic predictions of the LTT model in their shapes and amplitudes (Figure \ref{fig:pro_d}). 

The previous works which attempted to observationally test the LTT model using the sheet galaxies focused only 
on the anti-alignments between ${\bf j}$ and $\hat{\bf p}_{1}$ (or equivalently the tendency of ${\bf j}$ lying on the 
planes of the sheets) and measured the correlation strength between ${\bf j}$ and $\hat{\bf p}_{1}$ through fitting the 
observationally obtained $p(\vert j_{z}\vert)$ to an empirically constructed formula 
\citep[e.g.,][]{tru-etal06,SW09,var-etal12}.
However, the detection of a signal of the ${\bf j}$-$\hat{\bf p}_{1}$ anti-alignment and the agreement of the observational 
result with the fitting formula do not suffice to declare a discovery of the linear tidal vestige, no matter how strong the signal is 
and no matter how well the agreement turns out to be. This is because the other nonlinear processes could contribute to the 
presence of strong ${\bf j}$-$\hat{\bf p}_{1}$ anti-alignment \citep[e.g.,][]{LP08,cod-etal12,dub-etal14,wel-etal14,lai-etal15}. 
Besides, the LTT model does not predict an arbitrary high signal of the ${\bf j}$-$\hat{\bf p}_{1}$ anti-alignment but 
rather puts an upper limit on the strength of the signal (or equivalently, on the possible values of the tidal correlation 
parameter $c_{t}$). 
In the current work,  even though we have derived the three probability density functions, 
$p(\vert j_{x}\vert),\ p(\vert j_{y}\vert),\ p(\vert j_{z}\vert)$, purely from the linear physics, without resorting to any empirical 
formula or to fitting procedure, we still witness good quantitative agreements of them with the observational results 
from the WMs galaxies. 

Another merit of our analysis is that we have been able to determine the eigenvalues and eigenvectors of the 
linear tidal tensor at the location of the WM sheet without resorting to a noisy reconstruction of the whole tidal field, 
owing to the particularly regular parallelepiped-like shape of the WM sheet. In the previous approaches to the detection 
of the large-scale tidal effects on the directions of the galaxy angular momenta \citep[e.g.,][]{LE07,zha-etal15}, 
the whole tidal field had to be reconstructed, which could be, however, quite noisy, depending on the background 
cosmology, and introducing an additional free parameter, the smoothing scale. 

Before drawing a conclusion, however, we would like to discuss two caveats. 
First, we have inferred the eigenvalues of the large-scale tidal field surrounding the WM sheet from the 
measurements of its three side lengths ($L_{1},\ L_{2},\ L_{3}$) under the assumption that the spatial extent  
of the distribution of the WMs galaxies observed in redshift space is at least proportional to that of the underlying dark 
matter in real space. 
Although recent numerical simulations showed that the sheets are in the quasi-linear regime containing very few  
nonlinear objects\citep{che-etal16,cau-etal14}, this assumption has to be justified with a hydrodynamic simulation. 

Second, we have implicitly assumed that the spin axis directions of the WMs galaxies are perfectly aligned with 
those of the underlying dark matter halos, when we adopted the conventional methodology to measure the spin directions 
of the WMs galaxies. The validity of this assumption, however, is rather limited given the results of the recent hydrodynamic 
simulations that the spin axes of the stellar parts of the galaxies are aligned not with those of the entire galactic halos but with 
those of the core parts of the halos \citep[e.g.,][]{hah-etal10,var-etal12,ten-etal14,ZS17}. 
Yet, these hydrodynamic results do not necessarily indicate that the observed alignments of the galaxy spin axes have no 
tidal origin. In fact,  the angular momenta of the stellar parts of the galaxies could develop even stronger correlations 
with the linear tidal field and may preserve better the initial tidal memory since the spin axes of the inner regions of the halos 
should be least affected by the external nonlinear processes and thus may not undergo severe modifications in the subsequent 
evolution.

Our future work is in the direction of investigating the above issues with a hydrodynamic simulations:  
how the ratios of the side lengths of the luminous parts of the sheets are related to those of the underlying dark matter 
distribution; how strongly the spin axes of the stellar parts of the galaxies are aligned with the linear tidal field compared with 
their dark halo counterparts. With these caveats, it is concluded that our result supports compellingly the LTT model and that 
the vestige of the linear tidal influences on the galaxy angular momenta are best preserved in the sheet environments. 

\acknowledgements

We are grateful to an anonymous referee for very useful comments. 
This work was supported by a research grant from the National Research Foundation (NRF) of Korea to the Center for  
Galaxy Evolution Research (No.2017R1A5A1070354). J.L. was also partially supported by the Basic Science Research 
Program through the NRF of Korea funded by the Ministry of Education (NO. 2016R1D1A1A09918491).  
S.C.R was also partially supported by the Basic Science Research Program through the NRF of Korea funded by the 
Ministry of Education (No. 2015R1A2A2A01006828 and 2018R1A2B2006445).


\clearpage
\clearpage
\clearpage
\clearpage
\clearpage
\clearpage
\clearpage
\clearpage
\begin{deluxetable}{ccc}
\tablewidth{0pt}
\setlength{\tabcolsep}{5mm}
\tablecaption{Spatial extents of the WMs galaxies}
\tablehead{$x_{sg}$ & $y_{sg}$ & $z_{sg}$\\
(Mpc) & (Mpc) & (Mpc) }
\startdata
$[-12.88,\ -1.66]$ & $[16.03,\ 24.99]$ & $[-3.10,\ -1.77]$ 
\enddata
\label{tab:wm}
\end{deluxetable}
\clearpage
\begin{deluxetable}{cccc}
\tablewidth{0pt}
\setlength{\tabcolsep}{5mm}
\tablecaption{Tidal link parameters and their errors.}
\tablehead{constraint & $N_{r}$ & $\bar{c}_{t}$ & $\sigma_{c_{t}}$ }
\startdata
none & $408$ & $0.24$ & $0.07$  \\
$d_{x}\ge d_{x,c}$ &$206$ & $0.21$ & $0.10$ \\
$d_{x}<d_{x,c}$ & $202$ & $0.27$ & $0.11$ \\
$d \ge d_{c}$ & $274$ & $0.29$ & $0.09$ 
\enddata
\label{tab:ct}
\end{deluxetable}


\clearpage
\begin{thebibliography}{000}
\bibitem[Arag{\'o}n-Calvo et al.(2007)]{ara-etal07} 
Arag{\'o}n-Calvo, M.~A., van de Weygaert, R., Jones, B.~J.~T., \& van der Hulst, J.~M.\ 2007, \apjl, 655, L5
\bibitem[Cautun et al.(2013)]{cau-etal13} 
Cautun, M., van de Weygaert, R., \& Jones, B.~J.~T.\ 2013, \mnras, 429, 1286
\bibitem[Cautun et al.(2014)]{cau-etal14} 
Cautun, M., van de Weygaert, R., Jones, B.~J.~T., \& Frenk, C.~S.\ 2014, \mnras, 441, 2923
\bibitem[Bett \& Frenk(2016)]{BF16} 
Bett, P.~E., \& Frenk, C.~S.\ 2016, \mnras, 461, 1338
\bibitem[Bland-Hawthorn \& Peebles(2006)]{BP06} 
Bland-Hawthorn, J., \& Peebles, P.~J.~E.\ 2006, Science, 313, 311
\bibitem[Bond et al.(1996)]{web96} 
Bond, J.~R., Kofman, L., \& Pogosyan, D.\ 1996, \nat, 380, 603
\bibitem[Brinckmann et al.(2016)]{bri-etal16} 
Brinckmann, T., Lindholmer, M., Hansen, S., \& Falco, M.\ 2016, JCAP, 4, 007
\bibitem[Catelan \& Theuns(1996)]{CT96} 
Catelan, P., \& Theuns, T.\ 1996, \mnras, 282, 436
\bibitem[Chen et al.(2016)]{che-etal16} 
Chen, S., Wang, H., Mo, H.~J., \& Shi, J.\ 2016, \apj, 825, 49
\bibitem[Codis et al.(2012)]{cod-etal12} 
Codis, S., Pichon, C., Devriendt, J., et al.\ 2012, \mnras, 427, 3320
\bibitem[Cuesta et al.(2008)]{cue-etal08} 
Cuesta, A.~J., Betancort-Rijo, J.~E., Gottl{\"o}ber, S., et al.\ 2008, \mnras, 385, 867
\bibitem[Doroshkevich(1970)]{dor70} 
Doroshkevich, A.~G.\ 1970, Astrofizika, 6, 581
\bibitem[Dubinski(1992)]{dub92} 
Dubinski, J.\ 1992, \apj, 401, 441 
\bibitem[Dubois et al.(2014)]{dub-etal14} 
Dubois, Y., Pichon, C., Welker, C., et al.\ 2014, \mnras, 444, 1453
\bibitem[Forero-Romero et al.(2009)]{for-etal09} 
Forero-Romero, J.~E., Hoffman, Y., Gottl{\"o}ber, S., Klypin, A., \& Yepes, G.\ 2009, \mnras, 396, 1815 
\bibitem[Forero-Romero et al.(2014)]{for-etal14} 
Forero-Romero, J.~E., Contreras, S., \& Padilla, N.\ 2014, \mnras, 443, 1090 
\bibitem[Geller \& Huchra(1989)]{great_wall} 
Geller, M.~J., \& Huchra, J.~P.\ 1989, Science, 246, 897
\bibitem[Gott \& Thuan(1978)]{GT78} 
Gott, J.~R., III, \& Thuan, T.~X.\ 1978, \apj, 223, 426 
\bibitem[Gott et al.(2005)]{sloan_wall} 
Gott, J.~R., III, Juri{\'c}, M., Schlegel, D., et al.\ 2005, \apj, 624, 463 
\bibitem[Hahn et al.(2007)]{hah-etal07} 
Hahn, O., Porciani, C., Carollo, C.~M., \& Dekel, A.\ 2007, \mnras, 375, 489
\bibitem[Hahn et al.(2010)]{hah-etal10} 
Hahn, O., Teyssier, R., \& Carollo, C.~M.\ 2010, \mnras, 405, 274
\bibitem[Haynes \& Giovanelli(1984)]{HG84} 
Haynes, M.~P., \& Giovanelli, R.\ 1984, \aj, 89, 758 
\bibitem[Helou(1984)]{hel84} 
Helou, G.\ 1984, \apj, 284, 471 
\bibitem[Hirv et al.(2017)]{hir-etal17} 
Hirv, A., Pelt, J., Saar, E., et al.\ 2017, \aap, 599, A31
\bibitem[Joachimi et al.(2015)]{align_overview} 
Joachimi, B., Cacciato, M., Kitching, T.~D., et al.\ 2015, \ssr, 193, 1 
\bibitem[Kim et al.(2016)]{kim-etal16}
Kim, S., Rey, S.-C., Bureau, M., et al.\ 2016, \apj, 833, 207
\bibitem[Laigle et al.(2015)]{lai-etal15} 
Laigle, C., Pichon, C., Codis, S., et al.\ 2015, \mnras, 446, 2744 
\bibitem[Lee \& Pen(2000)]{LP00} 
Lee, J., \& Pen, U.-L.\ 2000, \apjl, 532, L5
\bibitem[Lee \& Pen(2001)]{LP01} 
Lee, J., \& Pen, U.-L.\ 2001, \apj, 555, 106 
\bibitem[Lee(2004)]{lee04} 
Lee, J.\ 2004, \apjl, 614, L1 
\bibitem[Lee \& Erdogdu(2007)]{LE07} 
Lee, J., \& Erdogdu, P.\ 2007, \apj, 671, 1248 
\bibitem[Lee \& Pen(2008)]{LP08} 
Lee, J., \& Pen, U.-L.\ 2008, \apj, 681, 798-805 
\bibitem[Lee(2011)]{lee11} 
Lee, J.\ 2011, \apj, 732, 99 
\bibitem[Libeskind et al.(2013)]{lib-etal13} 
Libeskind, N.~I., Hoffman, Y., Forero-Romero, J., et al.\ 2013, \mnras, 428, 2489 
\bibitem[Navarro et al.(2004)]{nav-etal04} 
Navarro, J.~F., Abadi, M.~G., \& Steinmetz, M.\ 2004, \apjl, 613, L41 
\bibitem[Peebles(1969)]{pee69} 
Peebles, P.~J.~E.\ 1969, \apj, 155, 393 
\bibitem[Pen et al.(2000)]{pen-etal00} 
Pen, U.-L., Lee, J., \& Seljak, U.\ 2000, \apjl, 543, L107 
\bibitem[Porciani et al.(2002)]{por-etal02} 
Porciani, C., Dekel, A., \& Hoffman, Y.\ 2002, \mnras, 332, 325 
\bibitem[Shandarin(1994)]{sha-etal94} 
Shandarin, S.~F.\ 1994, Physica D Nonlinear Phenomena, 77, 342
\bibitem[Shandarin et al.(1995)]{sha-etal95} 
Shandarin, S.~F., Melott, A.~L., McDavitt, K., Pauls, J.~L., \& Tinker, J.\ 1995, Physical Review Letters, 75, 7
\bibitem[Slosar \& White(2009)]{SW09} 
Slosar, A., \& White, M.\ 2009, JCAP, 6, 009
\bibitem[Tempel et al.(2013)]{tem-etal13} 
Tempel, E., Stoica, R.~S., \& Saar, E.\ 2013, \mnras, 428, 1827
\bibitem[Tempel \& Libeskind(2013)]{TL13} 
Tempel, E., \& Libeskind, N.~I.\ 2013, \apjl, 775, L42 
\bibitem[Tenneti et al.(2014)]{ten-etal14} 
Tenneti, A., Mandelbaum, R., Di Matteo, T., Feng, Y., \& Khandai, N.\ 2014, \mnras, 441, 470
\bibitem[Trujillo et al.(2006)]{tru-etal06} 
Trujillo, I., Carretero, C., \& Patiri, S.~G.\ 2006, \apjl, 640, L111
\bibitem[Varela et al.(2012)]{var-etal12} 
Varela, J., Betancort-Rijo, J., Trujillo, I., \& Ricciardelli, E.\ 2012, \apj, 744, 82  
\bibitem[Velliscig et al.(2015)]{vel-etal15} 
Velliscig, M., Cacciato, M., Schaye, J., et al.\ 2015, \mnras, 453, 721 
\bibitem[Welker et al.(2014)]{wel-etal14} 
Welker, C., Devriendt, J., Dubois, Y., Pichon, C., \& Peirani, S.\ 2014, \mnras, 445, L46 
\bibitem[West et al.(1995)]{wes-etal95} 
West, M.~J., Jones, C., \& Forman, W.\ 1995, \apjl, 451, L5 
\bibitem[White(1984)]{whi84} 
White, S.~D.~M.\ 1984, \apj, 286, 38 
\bibitem[Zel'dovich(1970)]{zel70} 
Zel'dovich, Y.~B.\ 1970, \aap, 5, 84 
\bibitem[Zhang et al.(2015)]{zha-etal15} 
Zhang, Y., Yang, X., Wang, H., et al.\ 2015, \apj, 798, 17 
\bibitem[Zjupa \& Springel(2017)]{ZS17} 
Zjupa, J., \& Springel, V.\ 2017, \mnras, 466, 1625
\end{thebibliography}
\end{document}